# SSELab: A Plug-In-Based Framework for Web-Based Project Portals

Christoph Herrmann, Thomas Kurpick, Bernhard Rumpe
*Software Engineering*
*RWTH Aachen University*
*Aachen, Germany*
*http://www.se-rwth.de/*

*Abstract*—**Tools are an essential part of every software engineering project. But the number of tools that are used in all phases of the software development life-cycle and their complexity is growing continually. Consequently, the setup and maintenance of current tool chains and development environments requires much effort and consumes a lot of time. One approach to counter this, is to employ web-based systems for development tasks, because centralized systems simplify the administration and the deployment of new features. But desktop IDEs play an important role in software development projects today, and will not be replaced entirely by web-based environments in the near future. Therefore, supporting a mixture of hosted tools and tools integrated into desktop IDEs is a sensible approach.**

**In this paper, we present the SSELab, a framework for web-based project portals that attempts to migrate more software development tools from desktop to server environments, but still allows their integration into modern desktop IDEs. It supports the deployment of tools as hosted services using plug-in systems on the server-side. Additionally, it provides access to these tools by a set of clients that can be used in different contexts, either from the command line, from within IDEs such as Eclipse, or from web pages. In the paper, we discuss the architecture and the extensibility of the SSELab framework. Furthermore, we share our experiences with creating an instance of the framework and integrating various tools for our own software development projects.**

## I. Introduction

The complexity and the amount of tools that are used in software engineering projects are still increasing massively. As the tool infrastructures become more and more complex, their installation, maintenance, extension, and re-use get time-consuming and cost-intensive.

With advances in the areas of server virtualization and cloud-computing, one trend to alleviate this problem is using web-based systems in all phases of the software development life-cycle [1]. The benefits of centralizing tool infrastructures as hosted services are a simplified administration and deployment of new features and tools. Another advantage is the possibility to integrate legacy tools into modern tool chains. This is essential in an industrial context, where legacy tools are prevalent [2].

The trend towards adopting server-based systems can also be observed in the area of desktop systems. For example, employing virtual machines for software development ensures that all developers work with a consistent environment and do not have to spend much time on its setup [3]. These virtual machines can be deployed on the developers' local computers, but can also be hosted on server clusters with the developers connecting to them by remote desktop sessions.

Another approach to utilize the capacity of server infrastructures are web-based platforms or IDEs such as Eclipse RAP, which attempt to substitute rich client applications with browser-based clients. Consequently, web-based IDEs require that most of the artifacts the developers work with are hosted on a server as well.

Overall, these trends show that in the future an increasing amount of software development tools will run on server machines, either completely or partially. Nevertheless, desktop IDEs are still relevant today and will probably not be replaced completely by web-based IDEs in the near future, but need to be regarded as an efficient means to develop software [1]. Therefore, utilizing a combination of hosted core infrastructure tools and IDE integration of tools with higher demands on interactivity is still a reasonable approach for current software development projects [4].

In this paper, we present the *SSELab*, an extensible application framework for web-based project portals that allows a plug-in-based integration of development tools as hosted services. The framework was designed with the goal to migrate more tools from the desktop to the server, to take advantage of the available processing power of current server infrastructures and to ease the administration of development tool chains. The contribution of this work is the definition of an architecture for a framework that allows the integration of tools on the server-side. Especially tools that have mainly been designed for desktop systems can be migrated to a server-based environment. Furthermore, the framework allows to access these hosted tools from modern IDEs, such as Eclipse, from command line applications, or even from web pages. In addition to the architecture, we present our experiences with creating an instance of the framework and list some requirements for the efficient integration of tools.

The remainder of this paper is organized as follows. In Section II, we give an overview of the architecture of the SSELab. Section III describes the extension points of the framework and Section IV presents our experiences with creating an instance of the framework. Section V discusses related work and Section VI concludes the paper.

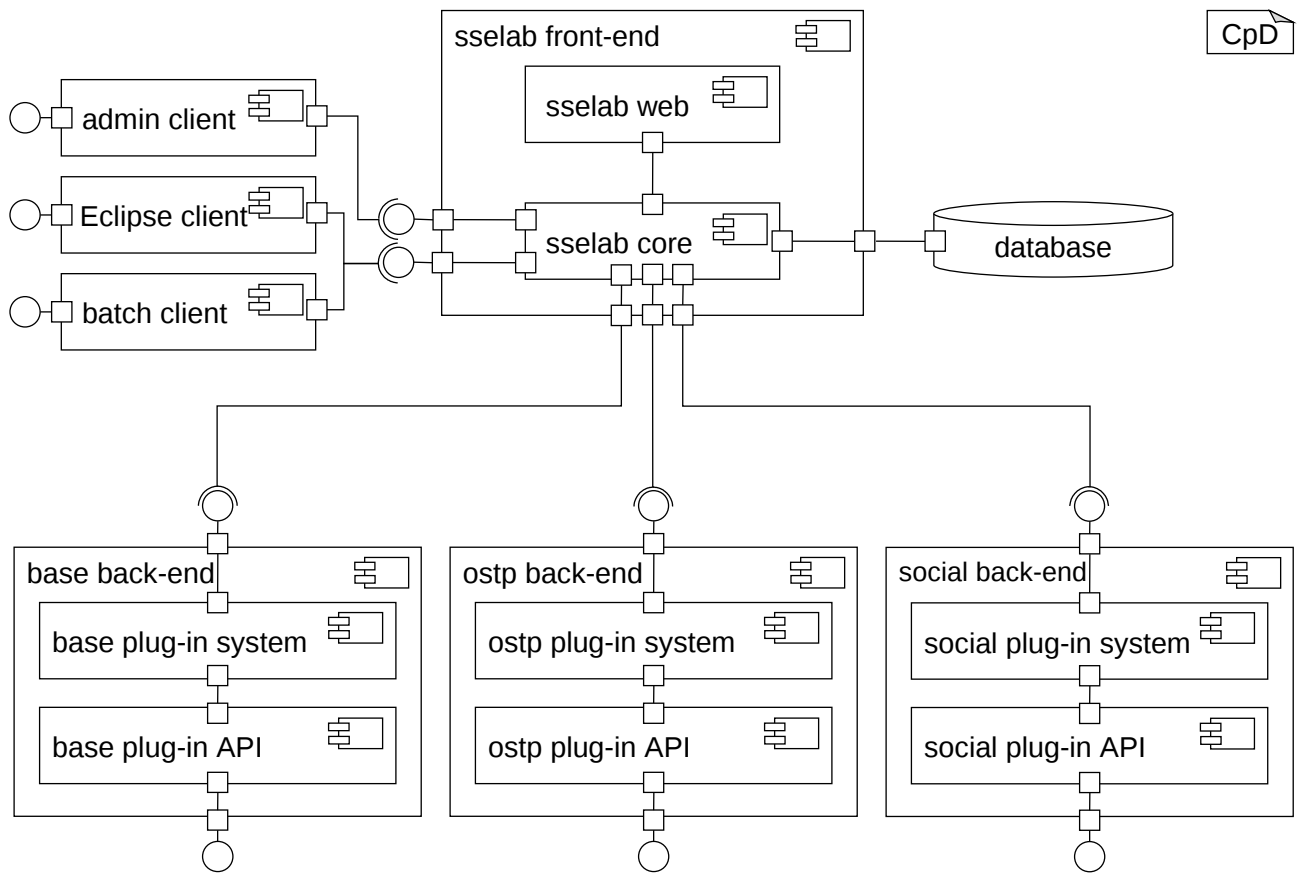

Figure 1. Architecture of the SSELab

## II. Architecture of the Framework

The SSELab has been designed to be both, an extensible framework and a ready-to-use web-based project portal. It consists of several components that are depicted in Figure 1, which shows the architecture in a UML component diagram (CpD). The central component of the SSELab is the *front-end*, which manages the different *back-ends* and the *clients* for the users and the administrators of the portal.

The *back-ends* are responsible for the configuration and execution of the integrated tools. In the current version of the SSELab three different types of back-ends are available. Each one supports tools from one of the categories described below. In the following discussion and the rest of this paper, these tools are also called *services* to emphasize the fact that they are integrated into the SSELab as hosted services.

- The *base back-end* allows the integration of web-based or server-based applications which are usually core infrastructure tools in software development projects. Therefore, they are called *base services*. Examples for tools in this category are revision control systems, bug-tracking tools, continuous integration servers, and wikis. These tools are mostly designed as server-based applications and are used as hosted services in development projects. Consequently, the characteristics of these tools are that they need a server infrastructure to run, usually provide their own user interface, and their own user management. They can either be used by web-browser or by stand-alone or IDE integrated clients.
- Tools that take a set of files as input and produce a set of output files can be integrated into the SSELab utilizing the *ostp*[1] *back-end*. Examples for such tools are code generators or file-based conversion tools. In most cases such tools are used locally as desktop or command line applications and possibly allow IDE integration. Another characteristic of these tools are the various parameterization options.
- The *social back-end* allows the users of the SSELab to connect to social networks and to import their profile information stored at these sites into their SSELab profile. The motivation for this category of services is to use social data to foster communication and team building. The availability of personal information in project portals, such as hobbies and interests, might help to build trust among co-workers and enhances the collaboration in distributed development projects [9].

All three back-end types have been designed based on the same concepts and consequently have a uniform architecture and runtime environment. They differ in the details of their implementation and the offered APIs for the development of services, which have been tailored towards each service category. A fundamental concept in the design of the back-ends is the utilization of plug-in systems. All back-ends use a plug-in system based on the OSGi specification [5] as runtime environment. Accordingly, the components of the back-ends shown in Figure 1, together with their runtime dependencies as well as the *service plug-ins* that integrate specific tools, are executed in an OSGi container. In addition, a Java web service stack and a HTTP server have been deployed in the container. This allows the back-ends to act as servers by publishing a web service interface that can be called by the front-end components. Furthermore, the mechanisms of the OSGi container allow the management of service plug-ins at runtime.

The *front-end* of the SSELab is the central web-based component for the administration of the system. It controls the interaction of all other components and manages the users, their projects, the back-ends with their installed service plug-ins, and the available clients. The features of the front-end can either be used by browser or by one of the clients, that communicate with it using the published web service interfaces. The front-end has been developed using the Java EE technology stack, it runs in an application server, and uses a database to store the relevant data. As previously mentioned, the communication of the front-end and the back-ends is also based on web service technologies. Therefore the front-end and the back-ends can be distributed among several servers and do not have to run on a single machine.

Most of the functionality of the SSELab is accessible through a web browser. But especially when working with files, using a browser is neither convenient nor efficient. Therefore, we developed different *clients* for the SSELab that can be used in various contexts. E.g., the client for administrative tasks allows the installation or the update of service plug-ins, and the clients for the users allow the execution of ostp services. There are different command line clients available as well as a client for Eclipse to demonstrate the integration into a modern IDE.

[1]Ostp used to be an acronym, but is no longer meant as one and will probably be replaced by a more meaningful name to identify file-based transformation tools.

## III. Extensibility of the Framework

Due to the framework approach, the extensibility of the SSELab is an essential concept that is reflected in the design of all components. The decoupling of the front-end from the back-ends and the clients, as well as the usage of plug-in systems, allows to extend the SSELab in several ways.

### A. Development of New Back-End Types

In the previous Section II, we have described the three supported back-end types. The creation of new back-end types is possible but currently requires an extension of the front-end components. The reason for this is that the services of each category are used in different contexts. For example, base services are used in the context of projects, i.e. they are configured for each project and all project participants get access to the services based on their specific role in the project. On the contrary, ostp services can be used independently of a specific project because their execution does not require storing state information between consecutive invocations. Social services are only used by each user individually in the context of a profile.

Consequently, extending the SSELab with a new back-end type requires knowledge of the front-end code and its extension to support the new service category in a user-friendly way. Simplifying the development of new back-end types is still an area of improvement, but not a pressing issue, because in our experience, new back-end types are not required on a regular basis.

### B. Deployment of Back-End Instances of Existing Types

The front-end supports an arbitrary number of instances of the three back-end types. At runtime, an administrator can register new instances with their unique URL in the front-end. The management of the back-ends is based on the concept that all instances of one specific type offer exactly the same web service interface. The front-end only uses these interfaces for the communication with the back-ends, if at least one back-end has been registered. As a result, the deployment of the front-end is decoupled from the deployment of the back-ends.

The management of the back-end instances at runtime requires mechanisms that ensure a consistent state among all instances. For this reason, the front-end stores a reference state for all back-end types including the installed service plug-ins with their dependencies. Therefore, the front-end can restore a consistent state, if necessary. This may either be relevant after a failure or misconfiguration of a back-end or if a new back-end instance has been deployed and needs to be initialized.

### C. Integration of Tools with Service Plug-Ins

The most important extension point of the SSELab is the possibility to integrate development tools with service plug-ins. Each tool that should be integrated in a specific version needs its own plug-in. But before the plug-in can be developed, the integration of the tool has to be prepared. The necessary steps for this vary depending on the service category as well as the features and interfaces of the tool.

- To integrate tools as base services, the required server infrastructure has to be installed, their configuration has to be automated, their user interface may need to be customized and the user management has to be adapted to use the security mechanisms of the SSELab.
- Most of tools in the ostp service category are designed for local usage. Their integration needs a server-based installation of the tool. The ostp back-end supports several ways to invoke a tool, e.g., by calling a Java API, if it is Java-based, by starting a process on the back-end host, or by calling a web service. Additionally, the parameterization of the tool needs to be mapped to the mechanisms of the SSELab.
- Social services integrate features of social networks. Therefore, the authentication of a user at the social network has to be supported as well as the automated fetching of profile information by API calls.

After the integration of a tool has been prepared, the service-plug-in can be developed. All back-ends of the SSELab define a Java-based framework that encapsulates all the OSGi-specific implementation. A developer only needs to subclass a class from the corresponding plug-in API and implement or override its methods. Afterwards, the code needs to be packaged in a jar-file, which can be uploaded by an administrator over the web pages of the front-end.

### D. Service-Specific Clients

The SSELab comes with a set of clients either for administrative tasks or for using the integrated services from the command line or from within Eclipse. The clients for administrative tasks are primarily designed to allow the automation of these tasks. An example for this is the scripted installation of a tool and its plug-in as part of a build or release process. The clients for using the services are primarily designed to be an alternative to browser-based clients in order to overcome the limitations when dealing with the input and output files of ostp services. Therefore, not all service categories are supported by these clients as discussed below.

- Base services can either be used by web browser or have many ready-to-use clients available. As the SSELab was designed to support existing clients, we have not seen the need to develop custom clients for base services.
- The clients of the SSELab for using services are primarily designed for ostp services, because these services integrate tools that are normally used locally.
- The social services are used in the context of user profiles. Using a web browser is common and efficient and sometimes even mandatory. Therefore, we have not created custom clients for this service category as well.

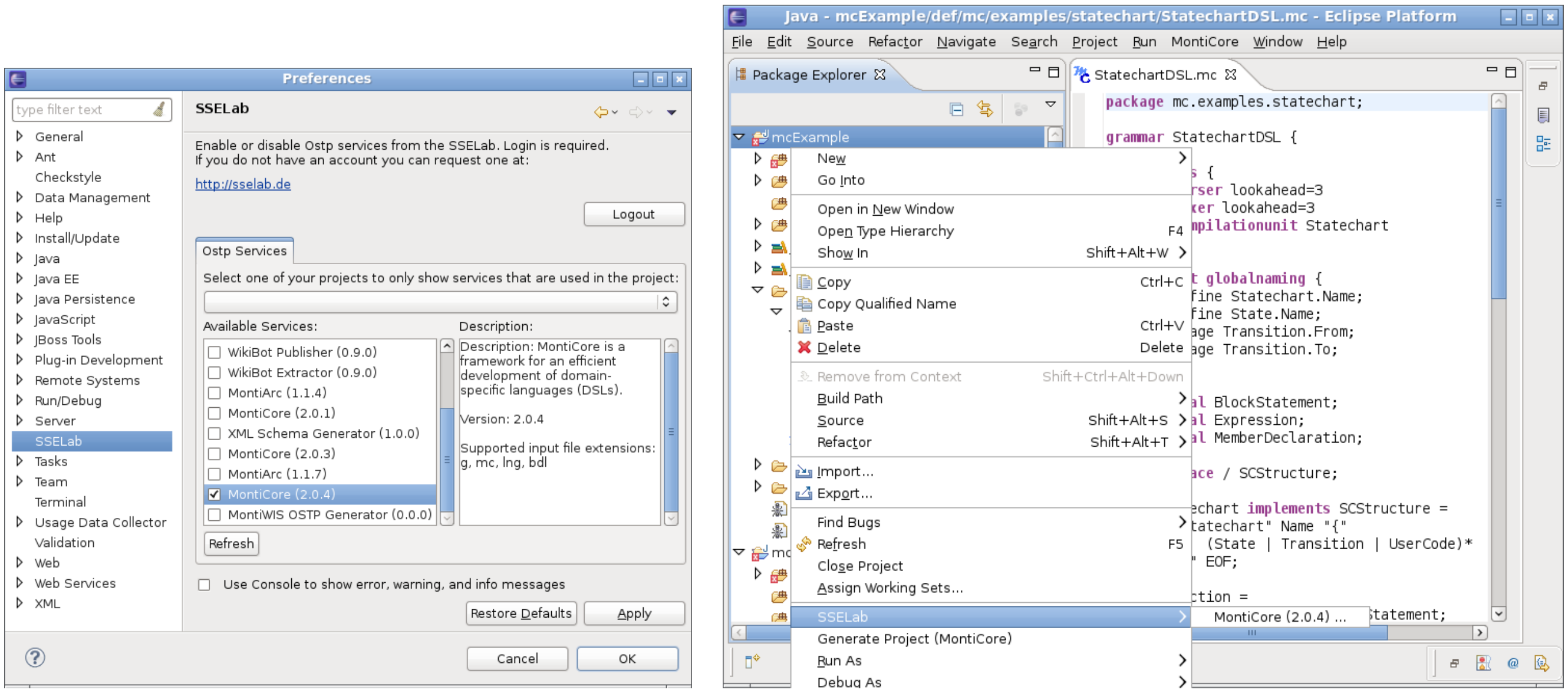

(a) SSELab preference page with available ostp services

(b) SSELab menu entry with MontiCore service

Figure 2. Generic Eclipse client for ostp services

In conformance with the fundamental concepts of the SSELab, the clients have been designed to be usable out-of-the-box in different contexts and to be extensible by service-specific clients. For example, the Eclipse client can be used to invoke any installed ostp service. Consequently, the interface of this client is generic in order to support all services. Figure 2 shows two screenshots of the client. Figure 2(a) depicts the preference page after the authentication of a user. All available ostp services are shown with their name, version, and description. From this list the user can select the services he wants to use. A service can then be invoked by clicking on a project, folder, or file in Eclipse and by selecting the corresponding menu entry. The menu is shown in Figure 2(b) with an entry for the MontiCore service. After the menu entry has been selected, a wizard appears that allows the user to specify parameter values, the output directory, and any other necessary data for the service invocation.

The Eclipse client can be used to execute all available ostp services. But its user interface does not include service-specific features and may not support a service in the most efficient way. Therefore, it can be extended by service-specific clients. In the following Section, we give an example for this based on the MontiCore tool.

## IV. Creating an Instance of the Framework

We have deployed an instance of the SSELab framework at http://sselab.de. It is used by our chair for teaching, research, and industry projects and by other chairs of our university. In the following, we give an overview of some services we have already integrated or are working on at the time of writing. Some of the currently available base services are summarized in the following list.

- *Subversion* and *Git* as revision control systems.
- *Trac* as project management and bug tracking tool.
- *MediaWiki* as wiki-engine.
- WebDAV-based *storage* service.
- *Feedback Service*, a service that allows customer feedback from within a developed product.
- *Jenkins* as continuous integration server.
- *Nexus* as repository manager.
- *Wordpress* as blogging tool.

Most of the following ostp services we have integrated, are tools that we develop at our chair as part of our research projects.

- *MontiCore*, a framework for the development of textual domain-specific languages (DSLs).
- *MontiArc*, a framework for modeling and simulation of distributed information flow architectures.
- *MontiWIS*, a framework for generation of web information systems using class and activity diagrams.
- *XML Schema Generator*, a service that generates a XML Schema Definition from a set of XML documents.
- *WikiBot*, a service that allows to download and to upload articles from and to wikis.

At the time of writing, the social back-end has just recently been developed. We have created plug-ins for the following social networks. This initial selection mostly has technical reasons, as these networks provide different APIs.

- *Google*
- *LinkedIn*
- *Myspace*

The SSELab instance we have deployed for our needs, mostly targets development projects with agile development processes using nightly builds, continuous integration, test metrics, etc. But the framework itself does not impose any specific development process, but can be tailored to support any development process by the integration and combination of services. In addition, we also have used this instance for other project types successfully, e.g. for project acquisition, to write books and theses.

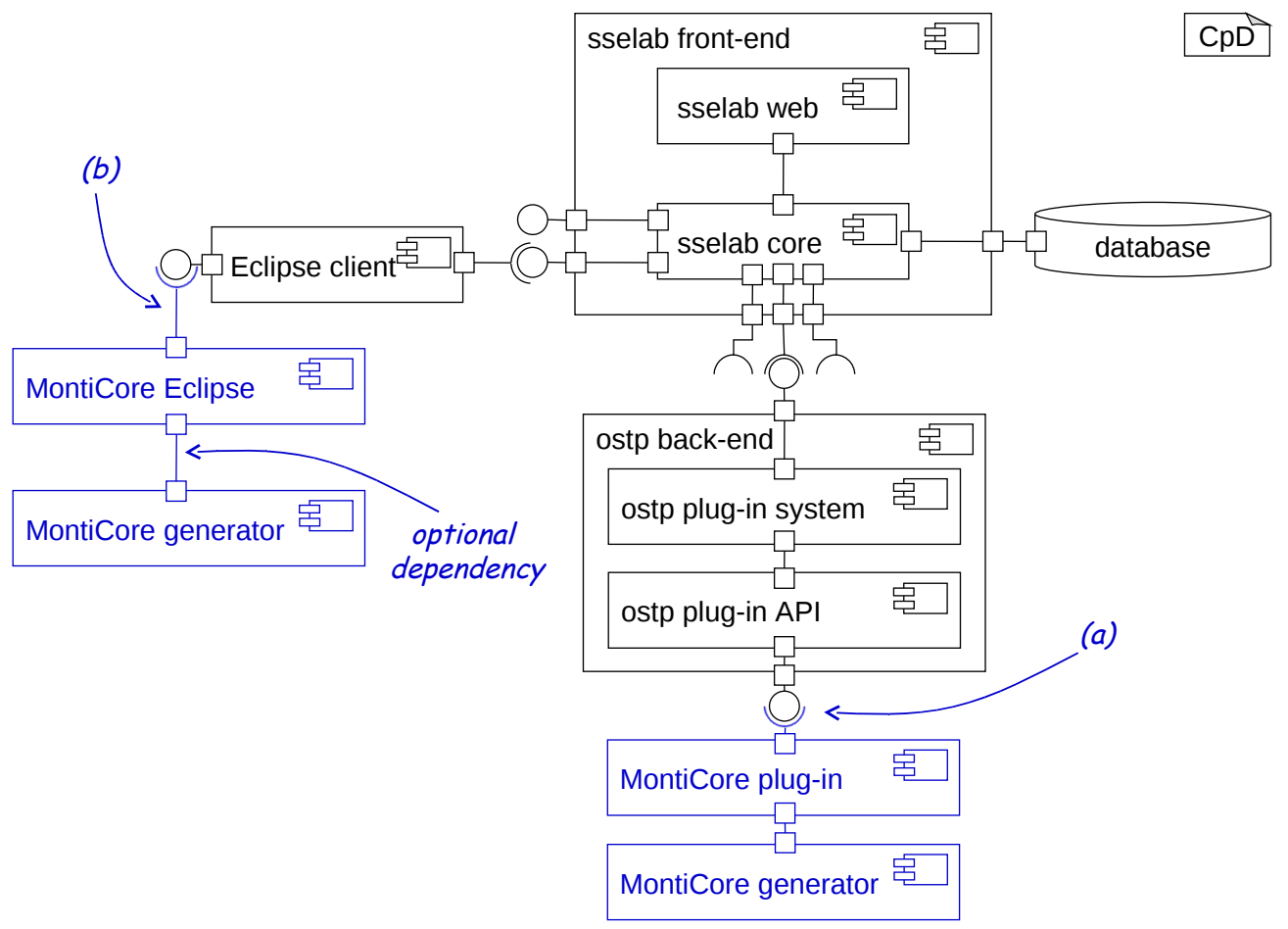

Figure 3. Integration of MontiCore into the SSELab

### A. Integration of MontiCore

In this section, we briefly present the integration of MontiCore [6] as ostp service. MontiCore itself has been created at our chair and is a framework for the development of domain-specific languages (DSLs). The DSL is defined by an extended grammar format and MontiCore generates components for processing the documents of the DSL. Examples for these components are parser, AST classes, symbol tables or pretty printers. Additionally MontiCore provides IDE integration for Eclipse.

Figure 3 shows the integration of MontiCore in the SSELab. Due to space restrictions only the relevant parts of the SSELab architecture and a minimal set of the actual MontiCore components are depicted. The MontiCore generator together with its runtime dependencies is installed in the ostp back-end. The MontiCore plug-in uses the ostp plug-in API to integrate the generator into the SSELab ((a) in Figure 3). The generic Eclipse client that has been described in Section III-D can be used to execute the generator. Additionally, MontiCore itself provides Eclipse integration and contributes an editor for its extended grammar format with syntax highlighting, an outline view and menu contributions to invoke the generator. The MontiCore Eclipse plug-in has an optional dependency to the MontiCore generator. If this dependency is available at runtime, MontiCore uses the local version. If it is not available, MontiCore automatically calls the ostp service to perform the code generation ((b) in Figure 3). This is handled in a transparent way, so that the users do not have to use a different workflow. The only difference is, that for the remote generation an authentication dialog might pop up if the user has not logged in to the SSELab.

### B. Requirements for the Integration of Tools

While creating the instance of the framework, integrating tools like MontiCore, and extending the framework itself, we have gathered experience with the integration of software development tools. In our experience, a lot of tools have been developed without giving enough attention to an integration in platforms like the SSELab. In the following, we share some of our experiences from integrating tools in the different categories and derive some requirements for a simpler integration. The steps to integrate a tool in each category have already been summarized in Section III-C.

Until now, setting up the server infrastructure for base services has not been an issue for the tools we integrated into the SSELab. Most of them run in well-established web, application and database servers. In contrast, the support for an automated configuration varies among the tools we integrated. For example, Subversion and Trac offer command line interfaces and text-based configuration files and thus can be scripted easily. At the time we integrated MediaWiki and Wordpress, they focused on a manual installation using a web form. Hence, automating the configuration of these tools was unnecessarily complex. Overall, we can derive the following requirements from our experiences for tools in this category.

- Rely on well-established servers.
- Provide scripts or APIs for an automated configuration.
- Support common authentication mechanisms to allow single sign-on.
- Provide configurable fine-granular access control for authorization.

Ostp services are designed for local use, so that their integration in a server-based environment is usually not taken into account. Our experience with the integration of tools in this category led us to the following requirements.

- Ship with a minimal set of runtime dependencies.
- Use a modular design to separate the user interface from the application logic.
- Provide a command line interface and allow the tool to run headless even if tool is usually used in a graphical environment.
- Build the tool with support for multi-thread environments.

As the social back-end is new, we have not gained as much experience with the integration of social networks as we have with the other categories. Some initial requirements for social platforms are given in the following list.

- Support common authentication protocols.
- Provide a documented API for fetching the data of an authenticated user.

## V. Related Work

Nowadays, many popular web-based project portals exist, such as SourceForge, GitHub, Google Code, Assembla, and JavaForge. Most of them offer core infrastructure tools as hosted services, but only limited integration mechanisms for IDEs [7]. Additionally, there is little information about their architecture available, as most of them are commercial, so that a comparison of their design with the architecture of the SSELab is hard to accomplish.

In [7] different project portals have been surveyed. IBM Jazz is mentioned as the only portal that supports IDE integration. Jazz is based on Eclipse – both, its server and the client – and incorporates collaborative tools into the IDE [8]. Consequently, Jazz is extensible and based on a plug-in system on the server-side, which is a fundamental concept the SSELab as well. There has also been work to integrate social network information into Jazz. In [9] the authors describe their work, in which they have added group awareness to Jazz in addition to the already supported presence and workspace awareness by integrating the FriendFeed aggregator service. Some potential limitations of Jazz have been mentioned in [10]. For example, Jazz imposes certain processes and if other processes are already in place this may be constraining. Also, it only works on the Eclipse platform which might be restricting and can bee seen as vendor lock-in, which is a risk when choosing a tool integration platform [11].

In [12] a collaborative platform called ProGET is described. This platform targets non-programmer researches and integrates several tools such as phpGroupWare, a mailing list manager, TWiki, WebDAV-based storage and a reporting tool. The platform has been created, taking the experiences from the development of PicoLibre (later renamed to PicoForge) into account. PicoLibre is a collaborative platform targeted at software engineering projects [13]. Both platforms have not been built as frameworks, but target projects in a specific domain, which has been identified as a limitation in [12]. While designing the SSELab, this has been taken into account. Therefore, the SSELab framework is not tailored towards a specific project type.

The authors of [14] discuss an extension of GForge called Davenport. They have extended GForge in order to replace its divers protocols (LDAP, SSH, FTP, etc.) with WebDAV/HTTP. Additionally, they have replaced CVS with Subversion and made a few other extensions to GForge. In the future work section of the paper, they state that the IDE integration of Davenport with Eclipse is a planned feature. Also, they say that collaborative development environments should have a pluggable architecture. But at that point in time this has not been a characteristic of the GForge architecture.

## VI. Conclusion

In this paper we introduced the SSELab as an extensible framework for web-based project portals. It allows the integration of tools for software development as hosted services using plug-in systems on the server-side. In order to allow the usage of the services as well as an integration into modern desktop IDEs, the SSELab provides a set of clients that can be used in different contexts. This architecture allows a flexible integration of development tools, whether they are web-based or desktop systems.